\documentclass[a4paper,11pt]{article}
\usepackage{pos}
\usepackage{subcaption}
\newcommand{\dd}{{\rm{d}}}
\newcommand{\nstep}{n_{\mbox{\tiny step}}}
\newcommand{\nev}{n_{\mbox{\tiny ev}}}
\newcommand{\nrelax}{n_{\mbox{\tiny relax}}}

\newcommand{\DKL}{\tilde{D}_{\mbox{\tiny{KL}}}}
\newcommand{\CP}{\mathrm{CP}}

\title{Out-of-equilibrium simulations to fight topological freezing}

\author[a]{Claudio Bonanno}
\author*[b]{Alessandro Nada}
\author[c]{Davide Vadacchino}

\affiliation[a]{Instituto de F\'isica T\'eorica UAM-CSIC, c/ Nicol\'as Cabrera 13-15, Universidad Aut\'onoma de
Madrid, \\ Cantoblanco, E-28049 Madrid, Spain}

\affiliation[b]{Dipartimento di Fisica,  Universit\'a degli Studi di Torino and INFN, Sezione di Torino, \\
  Via Pietro Giuria 1, I-10125 Turin, Italy}

\affiliation[c]{Centre for Mathematical Sciences, University of Plymouth, \\
Plymouth, PL4 8AA, United Kingdom}

\emailAdd{alessandro.nada@unito.it}

\abstract{Calculations of topological observables in lattice gauge theories with traditional Monte Carlo algorithms have long been known to be a difficult task, owing to the effects of long autocorrelations times. Several mitigation strategies have been put forward, including the use of open boundary conditions and methods such as parallel tempering. In this contribution we examine a new approach based on out-of-equilibrium Monte Carlo simulations. Starting from thermalized configurations with open boundary conditions on a line defect, periodic boundary conditions are gradually switched on. A sampling of topological observables is then shown to be possible with a specific reweighting-like technique inspired by Jarzynski's equality. We discuss the efficiency of this approach using results obtained for the 2-dimensional $\text{CP}^{N-1}$ models. Furthermore, we outline the implementation of our proposal in the context of Stochastic Normalizing Flows, as they share the same theoretical framework of the non-equilibrium transformations we perform, and can be thought of as their generalization.}

\FullConference{The 40th International Symposium on Lattice Field Theory (Lattice 2023)\\
July 31st - August 4th, 2023\\
Fermi National Accelerator Laboratory\\}

\begin{document}
\maketitle

\section{Introduction}

At very fine lattice spacing, the vacuum of Lattice regularized QCD with periodic
boundary conditions is well known to be characterized by the emergence of 
topological sectors. These are labeled by different values of the topological 
charge $Q$, and are separated by energy barriers whose height tends to infinity
as the continuum limit is approached. 
As the lattice spacing is reduced, standard 
Markov Chain Monte Carlo (MCMC) algorithms based on local updating algorithms
becomes less and less efficient in overcoming these barriers, and, eventually 
the Markov chain remains trapped in a fixed topological sector. 
This phenomenon, known as ``topological freezing'', causes topological quantities 
to suffer from very long autocorrelation times, that increase exponentially as the
continuum limit is approached, see Refs.~\cite{Alles:1996vn,deForcrand:1997yw,Lucini:2001ej,DelDebbio:2002xa,Leinweber:2003sj,
DelDebbio:2004xh,
Bonati:2017woi,Bonanno:2018xtd,Berni:2019bch,Bonanno:2020hht,Athenodorou:2021qvs,Bennett:2022ftz,Bonanno:2022yjr,Bonanno:2022hmz}.

A strategy to mitigate this issue is provided by Open Boundary Conditions (OBC) in the temporal direction, see Refs.~\cite{Luscher:2011kk,Luscher:2012av}, 
that effectively remove the barriers between topological sectors: 
in MCMC simulations with such boundary conditions, topological observables feature 
much smaller autocorrelation times. Yet, OBC induce sizeable 
finite-size effects, and relevant observables can be computed only 
on portions of the volume far enough 
from the boundaries. Moreover, a notion of global topological charge is lost. 
Another promising alternative proposed to mitigate topological freezing 
is known as Parallel Tempering on Boundary Conditions 
(PTBC), see Ref.~\cite{Hasenbusch:2017unr}: in this state-of-the-art approach, 
replicas with different boundary conditions, interpolating from open to periodic, 
are simulated simultaneously and configurations of neighbouring replicas 
are swapped using a Metropolis step. This allows for an efficient sampling of topological observables on the replica with periodic boundary conditions (PBC), by exploiting 
the relatively short autocorrelation time of the replica with OBC and bypassing complications introduced by OBC.

In this contribution, we propose a new MCMC method based on out-of-equilibrium 
evolutions inspired by Jarzynski's equality, see Ref.~\cite{Jarzynski:1996oqb}, a well 
known result in non-equilibrium statistical mechanics. 
This approach has been widely used in lattice field theory as well, namely in the computation of interface free energies, see Ref.~\cite{Caselle:2016wsw}, of the equation of state, see Ref.~\cite{Caselle:2018kap}, of the renormalized coupling of gauge theories, see Ref.~\cite{Francesconi:2020fgi}, and of the 
entanglement entropy of lattice field theories, see Ref.~\cite{Bulgarelli:2023ofi}. 
Moreover, it has also been combined with Normalizing Flows (see Ref.~\cite{papamakarios2021}), 
a deep-learning architecture that has been recently applied to lattice field theories:
see Ref.~\cite{Albergo:2021vyo} for an introduction.
In this new framework, called Stochastic Normalizing Flows (SNFs), see Refs.~\cite{wu2020stochastic,Caselle:2022acb},
MCMC updates that compose out-of-equilibrium evolutions are combined with discrete coupling 
layers, i.e. the building blocks that compose Normalizing Flows, resulting in an 
improvement of the purely stochastic approach.

In the context of topological freezing mitigation, 
out-of-equilibrium evolutions can leverage the advantages of 
OBC--small autocorrelation times--while avoiding its pitfalls--the 
complication introduced by the boundaries--by a direct sampling of the PBC 
theory via an appropriate reweighting-like technique. In the following, we test 
this method on the $\CP^{N-1}$ models in two dimensions, and perform a direct 
comparison with results obtained in the same setting using the PTBC.

\section{Out-of-equilibrium evolutions}

Consider a family of actions 
$S_{c(n)}$ for a system of fields $\phi$, interpolating in $\nstep$ steps 
between a prior action
$S_0=S_{c(0)}$ and a target action $S=S_{c(\nstep)}$, where the \emph{protocol} $c(n)$
describes the value of one or more parameters in the action along the interpolation.
It is well known that the ratio between the prior and target partition functions, 
$Z_0$ and $Z$, can be calculated using Jarzynski's equality, see Ref.~\cite{Jarzynski:1996oqb},
\begin{equation}
 \frac{Z}{Z_0} = \langle \exp \left( -W \right) \rangle_{\mbox{\tiny f}},
\end{equation}
where $W$ is the \emph{generalized work}, defined as
\begin{equation}\label{eq:work}
 W (\phi_0, \phi_1, \dots, \phi) = \sum_{n=0}^{\nstep-1} 
 \left\{ S_{c(n+1)}\left[\phi_n\right] - S_{c(n)}\left[\phi_n\right] \right\}~.
\end{equation}
The generalized work is the change in the action of the system along a given protocol $c(n)$. 
The averaging operation $\langle \dots \rangle_{\mbox{\tiny f}}$ 
is defined to act on a quantity $A$ as follows,
\begin{equation}\label{eq:evolution_average}
\langle A \rangle_f = \int \dd \phi_0  \, \dd \phi_1 \dots \dd \phi \, q_0(\phi_0) \, P_{\mbox{\tiny{f}}}[\phi_0,\phi_1,\dots, \phi] \, A(\phi_0, \phi_1, \dots, \phi),
\end{equation}
where $P_{\mbox{\tiny{f}}} [\phi_0,\phi_1,\dots, \phi] = \prod_{n=0}^{N-1} P_{c(n)} (\phi_n \to \phi_{n+1})$ is the product of transition probabilities $P_{c(n)}$, each defined by the protocol $c(n)$.
The average in Eq.~(\ref{eq:evolution_average}) defines 
an \textit{out-of-equilibrium evolution}, which can be used to 
sample any observable $\mathcal{O}$ on the target probability distribution 
using a reweighting-like formula:
%
%
%
%
%
\begin{equation}
\label{eq:obs}
 \langle \mathcal{O} \rangle = \frac{\langle \mathcal{O}(\phi) \exp(-W((\phi_0, \phi_1, \dots, \phi))) \rangle_{\mbox{\tiny{f}}} }{\langle \exp(-W((\phi_0, \phi_1, \dots, \phi))) \rangle_{\mbox{\tiny{f}}} }.
\end{equation}

\noindent
In practical terms, an expectation value from Eq.~\eqref{eq:evolution_average} can
be computed as follows:
\begin{enumerate}
 \item Sample from the prior distribution $q_0=e^{-S_0} / Z_0$ 
        (e.g., a thermalized Markov Chain) the starting configuration $\phi_0$;
 \item Change the protocol parameter from $c(0)$ to $c(1)$ to compute 
        the first term of the sum of Eq.~\eqref{eq:work};
 \item Using a suitable MCMC algorithm, with transition 
        probability $P_{c(1)} (\phi_0 \to \phi_{1})$), generate the 
        configuration $\phi_1$, now not necessarily at equilibrium anymore;
 \item Repeat until the final value of the protocol $c(\nstep)$ has been reached.
        Once the final configuration $\phi$ has been generated, the expectation value
        of an observable $\mathcal{O}(\phi)$ can be computed according 
        to Eq.~\eqref{eq:obs}.
\end{enumerate}
In order to sample the space of intermediate configurations $\phi_0$, $\phi_1$, \ldots, 
this procedure is repeated $\nev$ times. In case the prior distribution is a 
thermalized Markov Chain, number of MCMC updates between successive starting 
configurations is also a relevant parameter, that we call $\nrelax$.


To assess the quality of the protocol chosen to perform the 
out-of-equilibrium evolutions, one can consider the Kullback--Leibler divergence 
between the forward and the backward evolutions,
\begin{equation}
\DKL(q_0 P_{\mbox{\tiny{f}}} \| p P_{\mbox{\tiny{r}}}) =  
\langle W \rangle_{\mbox{\tiny{f}}} + \log \frac{Z}{Z_0}  \geq 0~,
\end{equation}
where the inequality is a restatement of the Second Principle of Thermodynamics. 
The metric that we will use in the following is the 
Effective Sample Size (ESS), defined as:
\begin{equation}
 \mbox{ESS} = \frac{\langle \exp (-W) \rangle_{\mbox{\tiny{f}}}^2}{\langle \exp(-2W) \rangle_{\mbox{\tiny{f}}}} \in [0,1],
\end{equation}
which is equal to 1 in the case of a perfectly reversible evolution.

\section{Numerical results in the $2d$ $\CP^{N-1}$ models}

Our numerical experiments have been conducted on the two-dimensional $\CP^{N-1}$ models, 
as in the original PTBC study, see Ref.~\cite{Hasenbusch:2017unr}, 
employing the numerical setup described in Ref.~\cite{Berni:2019bch}, 
where PTBC is implemented on a system with
a tree-level $O(a)$ Symanzik-improved lattice action. 
More precisely, the family of lattice actions used along the 
out-of-equilibrium stochastic evolutions is defined by:
\begin{equation}\label{eq:action}
\begin{aligned}
S_{c(n)} = -2 N \beta_L\sum_{x,\mu}\left\{ k_\mu^{(n)}(x) c_1\Re\left[\bar{U}_\mu(x)\bar{z}(x+\hat{\mu})z(x)\right] \right. + \\
\left.
k_\mu^{(n)}(x+\hat{\mu})k_\mu^{(r)}(x)c_2\Re\left[\bar{U}_\mu(x+\hat{\mu})\bar{U}_\mu(x)\bar{z}(x+2\hat{\mu})z(x)\right]\right\} \, ,
\end{aligned}
\end{equation}
where $\beta_L$ is the inverse bare 't~Hooft coupling, $c_1=4/3$ and $c_2=-1/12$ 
are the Symanzik-improvement coefficients, and the factors $k_\mu^{(n)}(x)$ regulate 
the boundary conditions along a given defect of length $d$: namely, 
$k_\mu^{(n)}(x) \equiv c(n)$ for any site $x$ on the defect and $\mu=0$, 
while $k_\mu^{(n)}(x) \equiv 1$ otherwise. In practice this means that the boundary 
conditions on the defect at any given step $n$ of an out-of-equilibrium evolution 
follow the protocol $c(n)$. We always choose lattices with a physical volume of $V=(aL)^2$.

Our observable of choice is the topological susceptibility $\chi_t$:
\begin{equation}
\label{eq:chi}
 \chi_t = \frac{1}{V} \langle Q^2 \rangle,
\end{equation}
which we compute from the geometric definition of the lattice topological charge
\begin{equation}
Q= \frac{1}{2\pi} \sum_x \Im \, \log \Pi_{12}(x) \in \mathbb{Z} \, ,
\end{equation}
where $\Pi_{\mu \nu} (x) \equiv U_\mu (x) U_\nu (x+\hat{\mu}) \bar{U}_\mu (x+\hat{\nu}) \bar{U}_\nu (x)$.
In order to compute $\chi_t$ with PBC, we make use of Eq.~\eqref{eq:obs}: in particular, we consider evolutions that start from the probability distribution of a system with fully OBC along the defect of length $d$ (i.e., we set $c(0)=0$) and reach the probability distribution with PBC after $\nstep$ steps (i.e. $c(\nstep-1)=1$). 
We always use a protocol $c(n)$ that grows linearly with $n$, i.e., $c(n)=n/(\nstep-1)$. 

We performed simulations in various settings. Several values of the defect 
length $d$ were explored in the interval $[6,L]$, each defining a different 
prior system. The number of steps $\nstep$ separating the prior system 
from the target system was always chosen in the interval $[100,2000]$. 
Each non-equilibrium evolution started from a configuration belonging to
a thermalized ensemble of the prior system. These thermalized ensembles were
generated with a 1:4 mixture of local heat-bath and over-relaxation update algorithms, with two successive configurations 
being separated by either $\nrelax=110$ or $\nrelax=250$ full lattice sweeps.
The total number of out-of-equilibrium evolutions $\nev$ was tuned so that
the various simulations all have a comparable overall numerical cost. 
An overestimate of the latter is $(\nstep+\nrelax)\times \nev$. 
In order that a reliable comparison with the PTBC algorithm can be performed, 
we used the same simulation settings as a subset of those explored 
in Ref.~\cite{Berni:2019bch}, including the same MCMC updating procedures. For more
details, we refer to Ref.~\cite{Berni:2019bch}.

\begin{figure}[!t]
\centering
\begin{subfigure}{.5\textwidth}
\centering
 \includegraphics[scale=0.59,keepaspectratio=true]{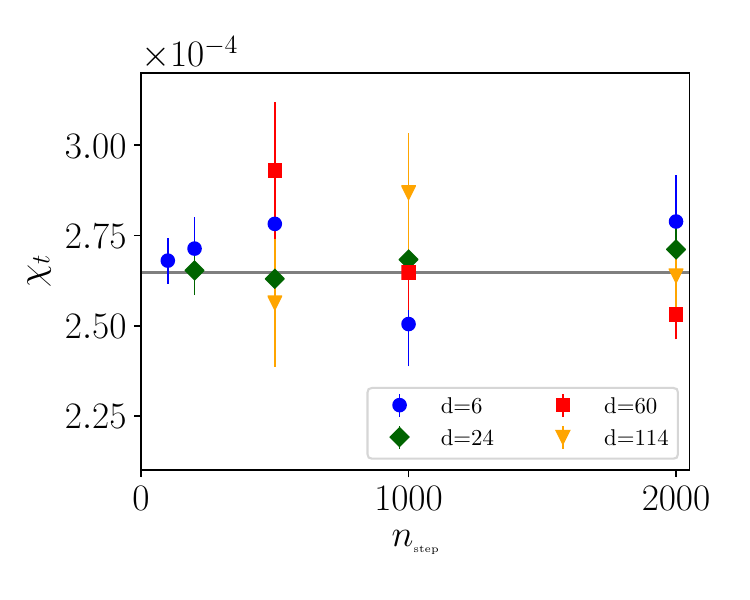}%
 \caption{$N=21$, $\beta_L=0.7$, $L=114$.}
 \label{fig:chi21}
\end{subfigure}%
\begin{subfigure}{.5\textwidth}
\centering
 \includegraphics[scale=0.59,keepaspectratio=true]{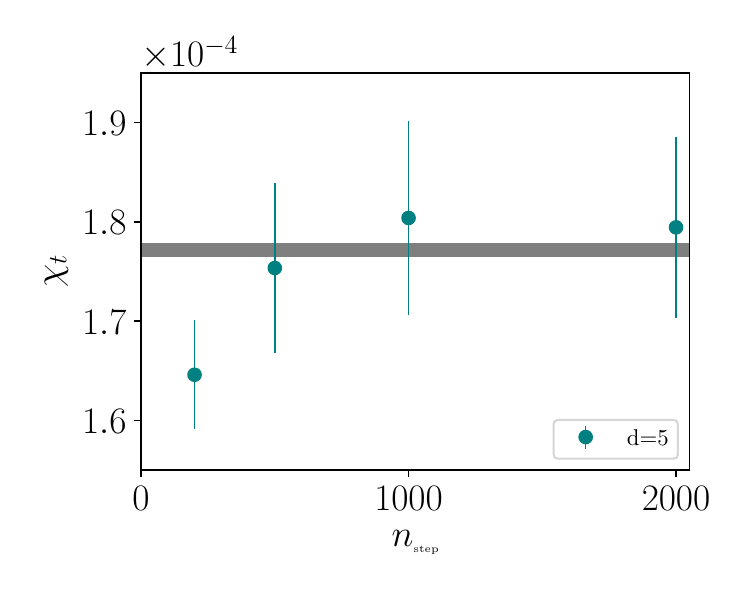}%
\caption{$N=41$, $\beta_L=0.65$, $L=132$.}
  \label{fig:chi41}
\end{subfigure}
\caption{Topological susceptibility for various values of $\nstep$ and $d$, compared to the result of ~\cite{Berni:2019bch} obtained with 11 (left) and 15 (right) replicas.}
\end{figure}

The first step in our analysis was to check that the value of $\chi_t$ 
obtained from the out-of-equilibrium evolutions is correct. That this is the 
case can be inferred from the results displayed
in Figs.~\ref{fig:chi21} and~\ref{fig:chi41}. A perfect agreement is found between
these values and the values obtained in Ref.~\cite{Berni:2019bch} with the PTBC algorithm.

\begin{figure}[!t]
\centering
\begin{subfigure}{.5\textwidth}
\centering
 \includegraphics[scale=0.59,keepaspectratio=true]{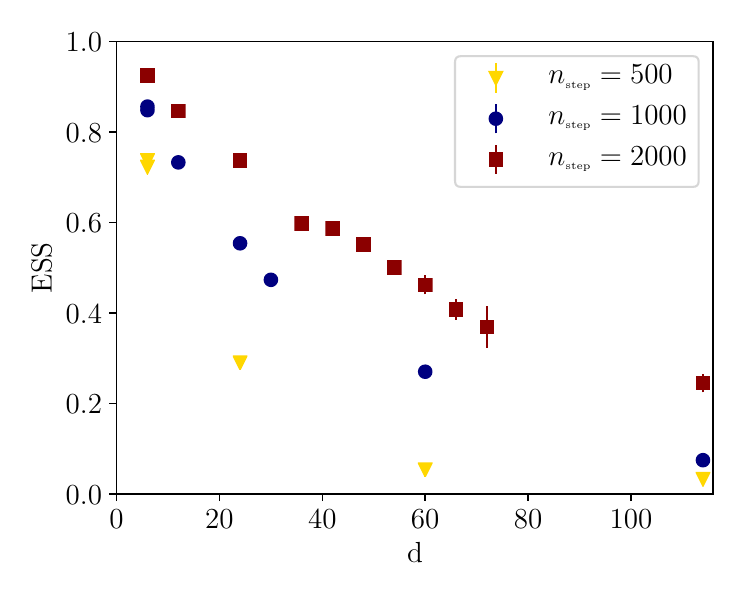}
 \caption{ESS as a function of the length $d$ of the defect.}
 \label{fig:ESS_defect}
\end{subfigure}%
\begin{subfigure}{.5\textwidth}
\centering
 \includegraphics[scale=0.59,keepaspectratio=true]{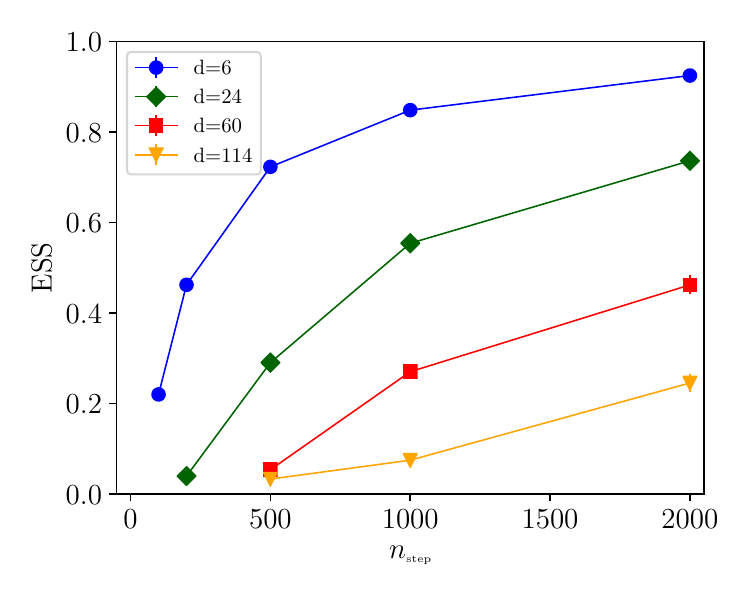}
 \caption{ESS as a function of the number of steps in each evolution.}
 \label{fig:ESS_nstep}
\end{subfigure}
\caption{ESS for different combinations of lengths of the defect and total number of steps in each evolution, for $N=21$, $\beta_L=0.7$ and $L=114$..}
\end{figure}

One of the goals of this study is to understand the magnitude of the 
numerical effort needed, in terms of 
out-of-equilibrium evolutions, in order to sample efficiently a system with PBC, 
starting from a system with a defect or with full OBC. To that aim, we display in 
Figs.~\ref{fig:ESS_defect} and~\ref{fig:ESS_nstep} the values of the 
Effective Sample Size as a function of $d$ (left panel) and $\nstep$ (right panel). 
Qualitatively speaking, a very small ESS (e.g., $< 0.05$) signals 
a (possibly extremely) inefficient sampling of the target distribution. 
Visual inspection of Fig.~\ref{fig:ESS_defect} shows that the ESS is a 
decreasing function of the defect length $d$. This is in agreement with the
fact that a prior system defined by larger defect length $d$ is farther 
from the target one with PBC. Thus, at a fixed value of $\nstep$, sampling
a system with PBC is increasingly more difficult as the prior system approaches
the full OBC (embodied by the choice $d=L$). At the same time, as can be seen
from Fig.~\ref{fig:ESS_nstep}, ESS is an increasing function of $\nstep$. Hence,
a simple solution to this issue seems to be to increase $\nstep$ to a sufficiently
large value for each $d$. Intuitively, this makes the evolution slower, as it is
close to a quasi-equilibrium evolution, and also more expensive to perform. 

\begin{figure}[!t]
\centering
\begin{subfigure}{.5\textwidth}
\centering
 \includegraphics[scale=0.59,keepaspectratio=true]{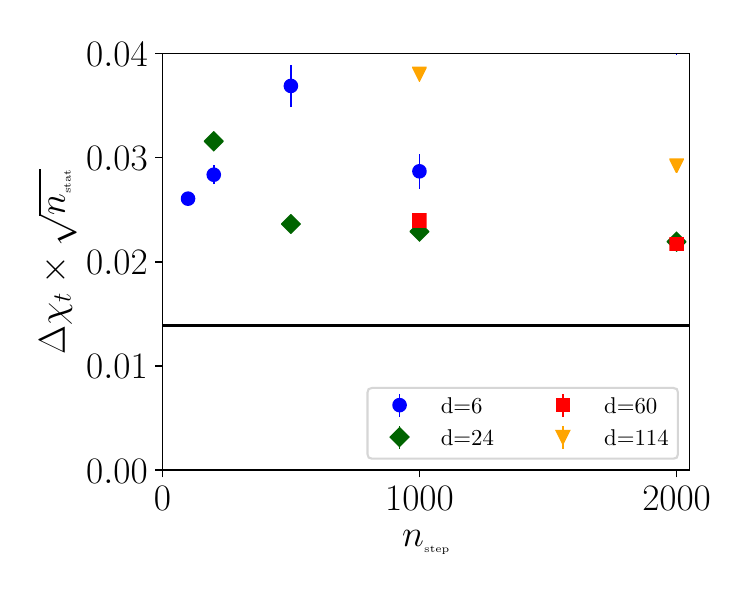}%
 \caption{$N=21$, $\beta_L=0.7$, $L=114$.}
 \label{fig:eff21}
\end{subfigure}%
\begin{subfigure}{.5\textwidth}
\centering
 \includegraphics[scale=0.59,keepaspectratio=true]{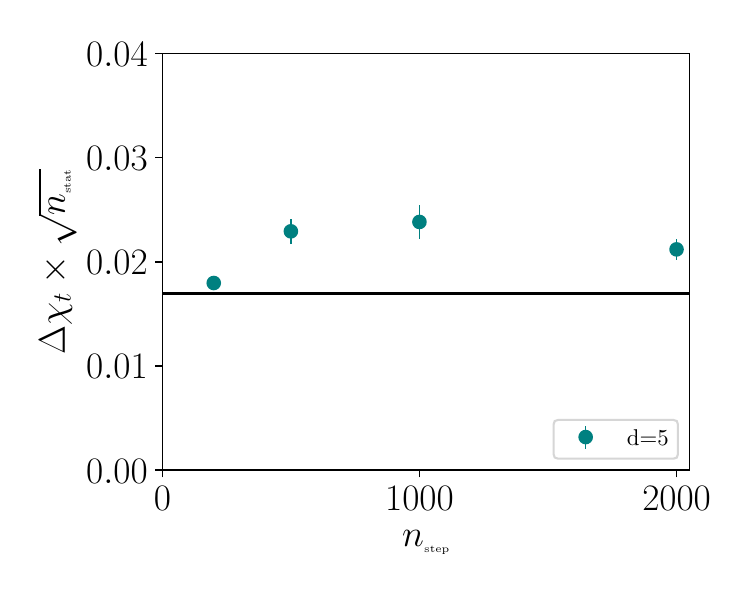}%
\caption{$N=41$, $\beta_L=0.65$, $L=132$.}
  \label{fig:eff41}
\end{subfigure}
\caption{Efficiency in the computation of $\chi_t$ for various values of $\nstep$ and $d$, compared to the parallel tempering (black band) from~\cite{Berni:2019bch}.}
\end{figure}

Finally, we display in Figs.~\ref{fig:eff21} and~\ref{fig:eff41} preliminary results 
concerning the efficiency of this method compared to the PTBC approach. We compare 
the error on the quantity $\chi_t$, multiplied by the square root of the total 
numerical effort spent to obtain the numerical results. 
In the case of the out-of-equilibrium evolutions, this is given by 
$(\nstep + \nrelax) \times \nev$, while in case of the PTBC algorithm this 
is given by the number of measurements multiplied by the number of replicas.
No specific choice in terms of the values of $\nstep$ or $d$ seems to be strikingly more
efficient than others. Moreover, it is quite encouraging to see that even for large 
values of $d$, non-equilibrium methods provide remarkably competitive results, although
the computational cost in terms of updates per evolution is larger. 
Generally speaking, non-equilibrium evolutions performed with $\nstep=1000$ or $2000$ 
enable a very precise sampling of the target distribution, even if the number of evolutions
$\nev$ themselves is comparably smaller. 
This is not surprising, as the use of relatively large values of $\nstep$ was essentially 
the same strategy already followed in the computation of the $\mathrm{SU}(3)$ 
equation of state in Ref.~\cite{Caselle:2018kap}.

In the case of $N=21$, displayed in Fig.~\ref{fig:eff21}, the preliminary results for out-of-equilibrium evolutions presented in this contribution are not yet comparable in efficiency with PTBC: we remark however that we opted for a very conservative estimation of the errors. In this respect, results obtained for $N=41$ shown in Fig.~\ref{fig:eff41} are even more encouraging, since autocorrelation times grow as a function of $N$, making topological freezing much worse at larger values of $N$.

\section{Conclusions and future outlooks}

In this contribution we showcased the first application of out-of-equilibrium methods
based on Jarzynski's equality towards the mitigation of so-called freezing of topological
observables in lattice field theory.

With this method, it is possible to leverage the milder autocorrelation times that 
enjoyed by lattice models with open boundary conditions while simultaneously
bypassing the complications they introduce. With an exact reweighting-like method, 
the physically-interesting observables with periodic boundaries could be computed, and
the preliminary numerical results on the $2d$ $\CP^{N-1}$ models show that this method 
is already competitive with state-of-the-art calculations performed with the PTBC 
algorithm, an approach that has recently seen wide use also for non-Abelian gauge 
theories in four dimensions.

An advantage of out-of-equilibrium evolutions over PTBC is that no additional replicas 
are needed, as each evolution is simulated independently. Moreover, as shown
by recent studies on Stochastic Normalizing Flows (SNFs), the combination of non-equilibrium
methods with the coupling layers of Normalizing Flows allows to improve their
efficiency even further. This opens up to a potentially exciting new development for the 
algorithmic approach pioneered in the present study, as a suitable training process of only 
moderate length could provide the values of the parameters of the coupling layers of SNFs.
In the future, we plan to explore this direction by implementing the above method with a 
suitable SNF architecture. This could boost even further its numerical efficiency, and 
provide a unique approach to mitigate topological freezing in 
the $2d$ $\CP^{N-1}$ models, and beyond.

\acknowledgments
The numerical simulations were run on machines of the Consorzio Interuniversitario per il Calcolo Automatico dell'Italia Nord Orientale (CINECA). A.~Nada acknowledges support by the Simons Foundation grant 994300 (Simons Collaboration on Confinement and QCD Strings) and from the SFT Scientific Initiative of INFN. The work of C.~Bonanno is supported by the Spanish Research Agency (Agencia Estatal de Investigación) through the grant IFT Centro de Excelencia Severo Ochoa CEX2020-001007-S and, partially, by grant PID2021-127526NB-I00, both funded by MCIN/AEI/10.13039/501100011033. The work of D.~Vadacchino is supported by STFC under under Consolidated Grant No.~ST/X000680/1.

\bibliographystyle{JHEP}
\bibliography{biblio}

\end{document}